\documentclass[12pt]{article}

\catcode`\@=11
\@addtoreset{equation}{section}

\global\arraycolsep=2pt
\usepackage{mathrsfs,amsbsy,amssymb,latexsym,amsfonts,amsmath,cite}
\usepackage{graphicx,color}

\newcommand\no{\nonumber}

\newcommand{\dd}{{\rm d}}

\allowdisplaybreaks

\begin{document}

\begin{flushright}
\parbox{4cm}
{KUNS-2660 \\ 
\today }
\end{flushright}

\vspace*{1.5cm}

\begin{center}
{\Large \bf 
Deformations of the Almheiri-Polchinski model}
\vspace*{1.5cm}\\
{\large Hideki Kyono\footnote{E-mail:~h\_kyono@gauge.scphys.kyoto-u.ac.jp}, 
Suguru Okumura\footnote{E-mail:~s.okumura@gauge.scphys.kyoto-u.ac.jp}
and Kentaroh Yoshida\footnote{E-mail:~kyoshida@gauge.scphys.kyoto-u.ac.jp}} 
\end{center}

\vspace*{0.5cm}

\begin{center}
{\it Department of Physics, Kyoto University, \\ 
Kitashirakawa Oiwake-cho, Kyoto 606-8502, Japan} 
\end{center}

\vspace{1cm}

\begin{abstract}

We study deformations of the Almheiri-Polchinski (AP) model 
by employing the Yang-Baxter deformation technique. 
The general deformed AdS$_2$ metric becomes a solution of a deformed AP model. 
In particular, the dilaton potential is deformed 
from a simple quadratic form to a hyperbolic function-type potential 
similarly to integrable deformations. 
A specific solution is a deformed black hole solution. 
Because the deformation makes the spacetime structure 
around the boundary change drastically and a new naked singularity appears,  
the holographic interpretation is far from trivial. The Hawking temperature is the same 
as the undeformed case but the Bekenstein-Hawking entropy is modified due to the deformation. 
This entropy can also be reproduced by evaluating the renormalized stress tensor 
with an appropriate counter term on the regularized screen close to the singularity.  
\end{abstract}

\setcounter{footnote}{0}
\setcounter{page}{0}
\thispagestyle{empty}

\newpage

\tableofcontents

\section{Introduction}

In the recent study of string theory, one of the most important issues is to understand 
a holographic principle \cite{Hooft, Susskind} at the full quantum level 
(For a review see \cite{Bousso}). 
The AdS/CFT correspondence \cite{M, GKP, Witten} is a realization of the holography. 
This is, however, a conjectured relation and there is no rigorous proof so far. 
The integrable structure behind the correspondence at the planar level 
has played an important role in checking conjectured relations in non-BPS regions 
(For a comprehensive review, see \cite{review}). 
But the proof is still far from the completion and furthermore 
it does not seem likely that the integrability would work in the presence of a black hole.  

\medskip 

Towards the complete understanding of holography, 
it is significant to try to construct a simple toy model of quantum holography. 
In fact, Kitaev proposed such a model \cite{Kitaev}, which is a variant of the Sachdev-Ye (SY) model 
\cite{SY}.  More concretely, this model is a one-dimensional system composed of $N \gg 1$ 
fermions with a random, all-to-all quartic interaction. 
This model is now called the Sachdev-Ye-Kitaev (SYK) model\footnote{For the recent progress 
on the SYK model, see \cite{PR,MS,GR,FGMS,Witten2,many,Klebanov,Nishinaka}. }.  
It should be remarked that the Lyapunov exponent computed from an out-of-time-order 
four-point function \cite{LO, Kitaev-corr} in the SYK model saturates the bound presented 
in \cite{MSS}. This is the onset to open a window to a toy model of holography 
because the Lyapunov exponent of black hole in Einstein gravity is $2\pi/\beta$ \cite{SS1,SS2}, 
where $\beta$ is the inverse of the Hawking temperature 
(For a black hole S-matrix approach, see \cite{P}.). 

\medskip 

A promising candidate of the gravity dual for the SYK model is a 1+1 D dilaton gravity. 
This system was originally introduced by Jackiw \cite{Jackiw} and Teitelboim \cite{Teitelboim} 
(For a nice reviw of the 1+1 D dilaton gravity system, see \cite{V}). 
From a renewed interest, the dilaton gravity with a certain dilaton potential 
was intensively studied in the recent work \cite{AP}, 
and this model is called the Almheiri-Polchinski (AP) model. A black hole solution exists 
as a vacuum solution of the AP model. They studied its various properties 
like the RG flow structure at zero temperature, the Bekenstein-Hawking entropy, the renormalized 
boundary stress tensor, and the contribution of conformal matter fields to the entropy. 
For the recent progress on the AP model, see \cite{MSY,EMV}.

\medskip 

In this paper, we are concerned with deformations of the AP model. 
Why is it so interesting to study the deformations?
There are some observation and motivation based on the recent progress. 
The first is an observation that the SY model is constructed 
by performing a disordered quench to an {\it isotropic} quantum Heisenberg magnet \cite{SY}. 
The Heisenberg model itself is integrable. Hence, supposing that the conjectured duality is true, 
it is natural to expect that integrable deformations of it lead to the associated deformations 
of the AP model.  
The second is a motive to understand the holographic duals of deformed AdS$_2$ geometries.   
Recently, a systematic way to perform integrable deformations, which is called the Yang-Baxter 
deformation \cite{Klimcik,DMV,MY-YBE}\footnote{For the (affine) symmetry algebras, 
see \cite{symmetry} and \cite{Sch-alg}).}, 
has been intensively 
studied\footnote{For Yang-Baxter deformations of type IIB superstring on 
AdS$_5\times$S$^5$\,, see \cite{DMV2,ABF,ABF2,HRT,LRT,HT-sol,scale,
KMY-Jordanian-typeIIB,LM-MY,MR-MY,Sch-MY,KMY-SUGRA,Stijn1,Stijn2,MY-duality,KKSY,KY-super,
HvT,ORSY,BW,OvT,HT-NA, BW-NA,Stijn3}}. 
However, the holographic duals of the deformed geometries 
have been poorly understood. In particular, even the location of the holographic screen has not been 
clarified, though there is a proposal \cite{Kameyama}. 
Hence it is important to get much deeper understanding of the simplest 
case like AdS$_2$\,. Furthermore, the Yang-Baxter deformation is not 
applicable to black hole geometries in general, because those cannot be described usually 
as a coset, homogeneous space. However, it is not the case for a 1+1 D black hole 
presented in this paper.
 
\medskip 

Based on the observation and motivation described above, 
we will study deformations of the AP model by employing the Yang-Baxter deformation technique. 
The general deformed AdS$_2$ metric 
becomes a solution of a deformed AP model. In particular, the dilaton potential is deformed 
from a simple quadratic form to a hyperbolic function-type potential 
similarly to integrable deformations. A specific solution is a deformed black hole solution. 
Because the deformation makes the spacetime structure 
around the boundary change drastically and a new naked singularity appears,  
the holographic interpretation is far from trivial. The Hawking temperature is the same 
as the undeformed case but the Bekenstein-Hawking entropy is modified due to the deformation. 
This entropy can also be reproduced by evaluating the renormalized stress tensor 
with an appropriate counter term on the regularized screen close to the singularity.  

\medskip 

This paper is organized as follows. 
In section 2 we study the most general Yang-Baxter deformation of AdS$_2$\,. 
Section 3 introduces the classical action of 1+1 D dilaton gravity 
and the AP model as a special case. 
In section 4 we study deformations of the AP model. 
The most general deformed metric constructed in section 2 
becomes a solution with a deformed dilaton potential. This deformed 
system allows a black hole solution as a specific solution like in the AP model. 
The Bekenstein-Hawking entropy is also computed. 
In section 5, we revisit the black hole entropy from the viewpoint of the renormalized 
boundary stress tensor. Putting a regularized screen close to a singularity, we 
evaluate the renormalized boundary stress tensor with an appropriate counter term. 
The resulting entropy nicely agrees with the Bekenstein-Hawking entropy 
computed in section 4. 
Section 6 is devoted to conclusion and discussion.

\section{Yang-Baxter deformations of AdS$_2$}

In this section, we consider the most general Yang-Baxter deformation of the AdS$_2$ metric. 
First of all, we briefly describe a coset construction of the Poincar\'e AdS$_2$  
Then we study the most general Yang-Baxter deformation of Poincar\'e AdS$_2$\,. 
As a result, we obtain a three-parameter family of deformed AdS$_2$ spaces. 

\subsection{Coset construction of AdS$_2$}

Let us recall a coset construction of the Poincar\'e AdS$_2$ metric (For the detail of the coset construction, for example, see \cite{BKLSY}).  

\medskip 

The starting point is that the AdS$_2$ geometry is represented by a coset 
\begin{equation}
\mbox{AdS}_2 ~~= ~~ SL(2)/U(1)\,. \label{coset}
\end{equation}
By using the coordinates $t$ and $z$\,, a coset representative $g$ is parametrized as 
\begin{equation}
g = {\rm exp}\left[ t H \right]
{\rm exp}\left[ (\log z) D \right]\,, 
\label{group para}
\end{equation}
where $H$ and $D$ are the time translation and dilatation generators, respectively. 
By involving the special conformal generator $C$\,, 
the $\mathfrak{sl}(2)$ algebra in the conformal basis is spanned as 
\begin{eqnarray}
[D,H] = H\,, \qquad [C,H] = 2 D\,, \qquad [D,C] = -C\,.
\end{eqnarray}
These generators can be represented by the $\mathfrak{so}(1,2)$ 
ones $T_I~~(I=0,1,2)$ like 
\begin{eqnarray}
H \equiv T_0 + T_2\,, \qquad C \equiv T_0 - T_2\,, \qquad 
D \equiv T_1\,,   
\end{eqnarray}
where $T_I$'s satisfy the commutation relations:
\begin{equation}
[T_0,T_1] = - T_2\,, \qquad  [T_1,T_2] = T_0\,, \qquad [T_2,T_0] = -T_1\,. 
\label{symmetric}
\end{equation}
In the following, we will work with $T_I$'s in the fundamental representation, 
\begin{eqnarray}
T_0=\frac{i}{2}\sigma_1\,,\qquad T_1=\frac{1}{2}\sigma_2\,,\qquad T_2=\frac{1}{2}\sigma_3\,, 
\end{eqnarray}
where $\sigma_i~(i=1,2,3)$ are the standard Pauli matrices. 

\medskip  

Note here that the coset (\ref{coset}) is symmetric as one can readily understand 
from (\ref{symmetric})\,. When vector spaces $\mathfrak{h}$ and $\mathfrak{m}$ are spanned 
as 
\[
\mathfrak{h} = \mbox{span}_{\mathbb{R}}\langle\, T_2\, \rangle\,, 
\qquad \mathfrak{m} = \mbox{span}_{\mathbb{R}}\langle T_0,T_1 \rangle\,,
\]
the $\mathbb{Z}_2$-grading structure is expressed as 
\begin{eqnarray}
[\mathfrak{h},\mathfrak{h}] \subset \mathfrak{h}\,, \qquad 
[\mathfrak{m},\mathfrak{h}] \subset \mathfrak{m}\,, \qquad 
[\mathfrak{m},\mathfrak{m}] \subset \mathfrak{h}\,. 
\end{eqnarray} 
When representing the $\mathfrak{sl}(2)$ algebra by  
a direct product (as vector spaces)\,, 
\[
\mathfrak{sl}(2) = \mathfrak{h} \oplus \mathfrak{m}\,, 
\]
the projection operator $P :~\mathfrak{sl}(2) \,\to\, \mathfrak{m}$ can be defined as 
\begin{eqnarray}
P(X) \equiv \frac{ {\rm Tr}(X~T_0)}{{\rm Tr} (T_0~ T_0) } T_0
+  \frac{ {\rm Tr}(X~T_1)}{{\rm Tr} (T_1~T_1) }T_1\,, \qquad X \in \mathfrak{sl}(2)\,. 
\label{projection}
\end{eqnarray} 

\medskip 

Now the Poincar\'e AdS$_2$ metric can be computed by performing coset construction. 
The left invariant one-form $J = g^{-1}\dd g$ is expanded as 
\[
J = e^{0}\, T_0 + e^1\, T_1 + \frac{1}{2}\omega^{01}\,T_2\,. 
\]
Here $e^0$ and $e^1$ are the zweibeins, and $\omega^{01}$ is the spin connection. 
With the parametrization (\ref{group para})\,, the zweibeins are given by 
\[
e^0 = \frac{\dd t} {z}\,, \qquad e^1 = \frac{\dd z}{z}\,. 
\]
By using the projection operator $P$ in (\ref{projection}) and the explicit expressions of 
the zweibeins $e^0$ and $e^1$\,, 
the resulting metric is obtained as 
\begin{eqnarray}
\dd s^2 &=& 2 \mbox{Tr} \left[ J P(J) \right]  = - e^0 e^0 + e^1 e^1\nonumber \\ 
&=& \frac{-\dd t^2 +\dd z^2}{z^2}\,. 
\label{AdS2}
\end{eqnarray}
This is nothing but the AdS$_2$ metric in the Poincar\'e coordinates. 

\medskip 

Hereafter, it is often convenient to use the the light-cone coordinates defined as 
\begin{equation}
x^{\pm} \equiv t \pm z\,. 
\label{lc}
\end{equation}
Then the metric is rewritten as 
\begin{eqnarray}
\dd s^2 = -{\rm e}^{2\omega (x^+,x^-)}\,\dd x^+ \dd x^- 
= -\frac{4 \dd x^+ \dd x^-}{(x^+-x^-)^2}\,. 
\end{eqnarray}
The exponential factor will play an important role in later discussion.

\subsection{The general Yang-Baxter deformation}

Let us next consider Yang-Baxter deformations of the AdS$_2$ metric (\ref{AdS2})\,. 
In the usual discussion, Yang-Baxter deformations \cite{Klimcik, DMV,MY-YBE} 
are performed for 2D non-linear sigma models. 
Then the anti-symmetric two-form is also involved as well as the metric. 
Here we will concentrate on the metric part only.  

\medskip 

The prescription of the deformation is very simple. It is just to insert a factor as follows: 
\begin{eqnarray}
\dd s^2 = 2 \mbox{Tr} \left[J\, \frac{1}{1-2\eta R_g \circ P}\, P(J)\right]\,. 
\label{YB-metric}
\end{eqnarray}
Here $\eta$ is a constant parameter which measures the deformation. 
Then $R_g$ is defined as a chain of operation like 
\begin{eqnarray}
R_g(X) \equiv g^{-1} \circ R(g X g^{-1}) \circ g\,, 
\end{eqnarray}
where $g$ is the group element in (\ref{group para})\,. 
The key ingredient is a linear operator $R:~\mathfrak{sl}(2) \to \mathfrak{sl}(2)$\,, 
and satisfy the (modified) classical Yang-Baxter equation [(m)CYBE]:
\begin{eqnarray}
[R(X),R(Y)]-R([R(X),Y]+[X,R(Y)])=c \cdot [X,Y] 
\qquad (X\,,Y\in \mathfrak{sl}(2))\,. \label{mCYBE}
\end{eqnarray}
Here $c$ is a real constant parameter. 
The case with $c \neq 0$ is the mCYBE and 
the case with $c=0$ is the homogeneous CYBE. 

\medskip 

We consider the most general deformations with the following $R$-operator
\begin{eqnarray}
R(T_I) &=& \widetilde{\Omega}_{IJ}M^{JK}\,T_K\,, \label{general}
\end{eqnarray}
where $\widetilde{\Omega}_{IJ}$ and $M^{IJ}$ are defined as\footnote{A similar study  
was done for the Nappi-Witten model \cite{NW} in \cite{KY-NW}.  
Note that the present definition of $M^{IJ}$ is slightly different  
from the one in \cite{KY-NW}.} 
\begin{eqnarray}
\widetilde{\Omega}_{IJ} &\equiv& \text{Tr}(T_IT_J)=\frac{1}{2}\eta_{IJ}\,,  \qquad 
M^{IJ} \equiv 
\begin{pmatrix}
0&~m_1~&~m_2\;\\
-m_1&0&m_3\\
-m_2&-m_3&0\\
\end{pmatrix}\,,
\label{ansatz}
\end{eqnarray}
Putting the ansatz (\ref{general}) into the (m)CYBE (\ref{mCYBE}) leads to an algebraic relation, 
\begin{eqnarray}
-m_1^2-m_2^2+m_3^2=4c\,.
\label{constraint}
\end{eqnarray}

\medskip 

After evaluating the expression (\ref{YB-metric}) with the general ansatz (\ref{general})\,, 
one can obtain the following metric: 
\begin{eqnarray}
\dd s^2 = \frac{-\dd t^2 + \dd z^2}{z^2-\eta^2\left(\alpha + \beta t+\gamma (-t^2+z^2) \right)^2}\,. 
\label{3-metric}
\end{eqnarray}
Here $\alpha$\,, $\beta$ and $\gamma$ are defined 
as linear combinations of $m_p~~(p=1,2,3)$ as follows: 
 \begin{eqnarray}
\alpha \equiv \frac{1}{2}(m_1+m_3)\,,\qquad 
\beta \equiv -m_2\,,\qquad 
\gamma \equiv \frac{1}{2}(m_1-m_3)\,.
\end{eqnarray}
When $\eta=0$\,, the undeformed metric (\ref{AdS2}) is reproduced. 
Note here that the four constant parameters $m_p~(p=1,2,3)$ and $c$ appear 
in our discussion. Then a constraint (\ref{constraint})\,, which comes from the (m)CYBE, 
is imposed. Hence, three of them are independent each other. 

\medskip 

The Ricci scalar curvature of the metric (\ref{3-metric}) is 
\begin{eqnarray}
\label{Ricci}
R=-2\left(1-\widetilde{\omega}\eta^2\right)\,
\frac{z^2+\eta^2\left(\alpha+\beta t +\gamma(-t^2+z^2)\right)^2}{
z^2-\eta^2\left(\alpha+\beta t +\gamma(-t^2+z^2)\right)^2}\,,
\end{eqnarray}
where we have introduced a new quantity, 
\begin{eqnarray}
\widetilde{\omega} &\equiv&  \beta^2+4\alpha\gamma=m_1^2+m_2^2-m_3^2 \nonumber \\ 
&=& -4c\,. 
\end{eqnarray}
At the last equality, the (m)CYBE (\ref{mCYBE}) has been utilized. 
The scalar curvature (\ref{Ricci}) changes (even its sign) depending on the values of parameters 
and coordinates, while it becomes a constant $-2$ in the undeformed limit $\eta \to 0$\,. 
The expression (\ref{Ricci}) indicates that the deformed geometry contains both AdS and dS 
in general.

\section{A brief review of the AP model}

In this section, we shall introduce the classical action of 1+1 D dilaton gravity 
system. Then we briefly describe the AP model and its properties related to our later discussion.

\subsection{1+1 D dilaton gravity system}

The dilaton gravity system in 1+1 dimensions is composed of the metric $g_{ab}~(a,b=0,1)$ 
and the dilaton $\Phi$\,. The coordinates are parametrized as $x^a = (x^0,x^1) = (t,z)$\,. 

\medskip 

The classical action $S$ is given by 
\begin{eqnarray}
\label{action}
&&S=S_{g,\Phi}+S_{\rm matter}\,,\no\\
&&S_{g,\Phi}=\frac{1}{16\pi G}\int\!d^2 x\, \sqrt{-g}\left(\Phi^2 R - U(\Phi)\right)\,,\no\\
&&S_{\rm matter}=\frac{1}{32\pi G}\int\!d^2 x\, \sqrt{-g}\, \Omega(\Phi)\, (\nabla f)^2\,.
\end{eqnarray}
Here $G$ is the Newton constant in 1+1 dimensions and $U(\Phi)$ is the dilaton potential. 

\medskip

In the following, we will work with the metric in the conformal gauge, 
\begin{eqnarray}
\dd s^2 &=& -{\rm e}^{2 \omega(x^+,x^-)}\dd x^+ \dd x^-\,, 
\end{eqnarray}
where the light-cone coordinates are defined in (\ref{lc})\,. 

\medskip

Then the equations of motion are given by 
\begin{eqnarray}
\partial_+(\Omega\partial_- f)+\partial_-(\Omega\partial_+ f)&=&0\,,\no\\
4\partial_+\partial_-\Phi^2 - {\rm e}^{2\omega}U(\Phi)&=&0\,,\no\\
2\partial_+({\rm e}^{-2\omega}\partial_- {\rm e}^{2\omega})
-\frac{1}{2}e^{2\omega}\partial_{\Phi^2}U(\Phi)&=&(\partial_{\Phi^2}\Omega)\partial_
+f \partial_-f\,,\no\\
- {\rm e}^{2\omega}\partial_+({\rm e}^{-2\omega}\partial_+\Phi^2)&=&
\frac{\Omega}{2}\partial_+f\partial_+f\,,\no\\
- {\rm e}^{2\omega}\partial_-({\rm e}^{-2\omega}\partial_-\Phi^2)&=&
\frac{\Omega}{2}\partial_-f\partial_-f\,.
\label{2D-eom}
\end{eqnarray}
The energy-momentum tenser for the matter field $f$ is normalized as 
\begin{eqnarray}
(T_{\rm matter})_{ab} &\equiv& 
-\frac{2}{\sqrt{-g}}\frac{\delta S_{\rm matter}}{\delta g^{ab}}\no\\
 &=&-\frac{\Omega(\Phi)}{16\pi G}\left(\partial_af \,\partial_b f -\frac{1}{2}g_{ab}\,
\partial^c f \partial_c f\right)\,.
\label{tm}
\end{eqnarray}
This expression (\ref{tm}) is valid for the general form of $\Omega(\Phi)$\,.

\subsection{The AP model} 

The AP model corresponds to a special case of 1+1 D dilaton gravity 
specified by the following condition:
\begin{eqnarray}
U (\Phi) = 2 - 2 \Phi^2, \qquad \Omega(\Phi)=1\,. \label{AP}
\end{eqnarray}
This model exhibits nice properties. Among them, we are concerned with 
the vacuum solution of this model. For our later convenience, we shall give a brief review 
of the work \cite{AP} by focusing upon the vacuum solution in the following. 

\medskip 

The general vacuum solution is given by 
\begin{eqnarray}
\dd s^2 &=&\frac{1}{z^2}(-\dd t^2 + \dd z^2) \,,\\ 
\Phi^2 &=& 1+\frac{a+b \,t +c \,(-t^2+z^2)}{z}\,, 
\label{dilaton-AP}
\end{eqnarray}
and depends on three real constants $a$\,, $b$ and $c$\,.  
This three-parameter family contains interesting solutions as specific examples. 
For example, the case with $a=1/2$\,, $b=0$ and $c=0$ corresponds to 
a renormalization group flow solution from a conformal Lifshitz spacetime  
to AdS$_2$ \cite{AP}, with an appropriate lift-up to higher dimensions. 

\begin{figure}[htbp]
\begin{center}
\includegraphics[scale=0.3]{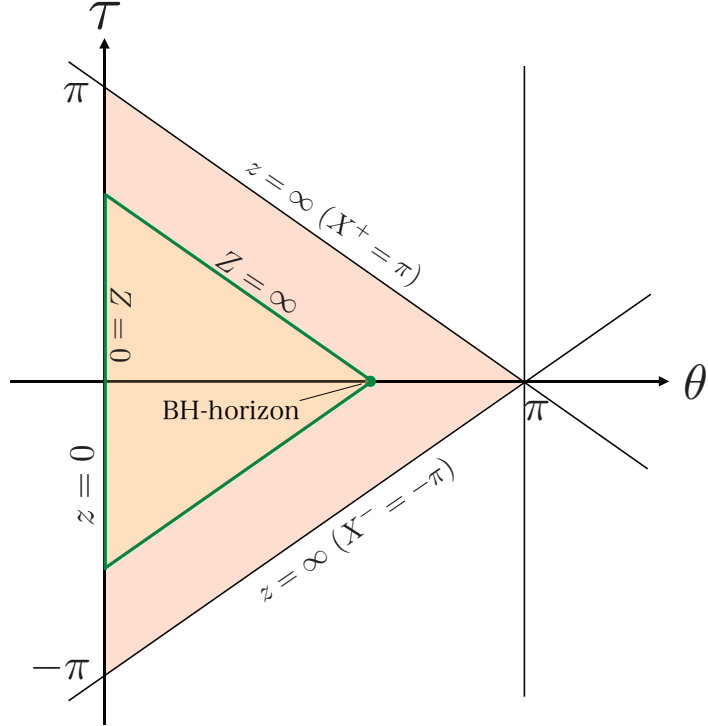}  
\vspace*{-0.5cm}
\caption{\label{fig:AdS penrose} \footnotesize Penrose diagram of the black hole solution. 
The global AdS$_2$ is parametrized by $\tau$ and $\theta$\,. The largest red triangle describes 
the Poincar\'e patch of AdS$_2$ as usual. The coordinate system (\ref{undeformed BH1}) covers 
the inside of smaller triangle bounded by green lines. The right vertex corresponds to 
the black hole horizon which is specified as the point that $T$ is finite but $Z$ is infinity.}
\end{center}
\end{figure}

\medskip 

Another intriguing example is a black hole solution specified with 
$a=1/2$\,, $b=0$ and $c=\mu/2$\,, where $\mu$ is a real positive constant. 
Then by performing a coordinate transformation, 
\begin{eqnarray}
x^\pm=\frac{1}{\sqrt{\mu}}
\tanh\left(\sqrt{\mu}\, (T\pm Z)\right),
\label{to-BH}
\end{eqnarray}
the solution is rewritten into the following form:
\begin{eqnarray}
\dd s^2 &=& \frac{4\, \mu}{\sin (2 \sqrt{\mu}\, Z)} (-\dd T^2+\dd Z^2)\,,\no\\
\Phi^2 &=& 1+ \sqrt{\mu}\,\text{coth}(2\sqrt{\mu}\, Z)\,. 
\label{undeformed BH1}
\end{eqnarray}
The new coordinates $T$ and $Z$ cover a smaller region which is in the inside of 
the entire Poincar\'e AdS$_2$\,, as depicted in Fig.\,\ref{fig:AdS penrose}. 

\medskip 

The background (\ref{undeformed BH1}) indeed describes a black hole geometry, 
but it may not be so manifest. 
To figure out the black hole geometry, it is nice to move to the Schwarzschild coordinates  
by performing a further coordinate transformation, 
\begin{eqnarray}
Z=\frac{1}{2\sqrt{\mu}}\text{arccoth}\left(\frac{\rho}{\sqrt{\mu}}\right)\,. 
\end{eqnarray}
Then the background (\ref{undeformed BH1}) can be rewritten as\footnote{The factor 4 is included so that 
the Bekenstein-Hawking entropy should match with the holographic computation. 
This normalization guarantees the matching of the bulk and boundary times (or temperatures). 
We are grateful to Ahmed Almheiri for this point. }  
\begin{eqnarray}
\dd s^2 = -4 (\rho^2-\mu)\dd t^2 + \frac{\dd \rho^2}{\rho^2-\mu}\,, \qquad 
\Phi^2 = 1+\rho\,. 
\end{eqnarray}
In this metric, the black hole horizon is located at $\rho=\sqrt{\mu}$\,,
and the Hawking temperature $T_{\rm H}$ can be evaluated in the standard manner as 
\begin{eqnarray}
T_{\rm H} =\left.\frac{1}{4\,\pi}\partial_\rho 
\sqrt{\frac{-g_{tt}}{g_{\rho\rho}}}\right|_{\rho=\sqrt{\mu}} =\frac{\sqrt{\mu}}{\pi}\,. 
\end{eqnarray}
Thus one can see that the background (\ref{undeformed BH1}) describes a black hole 
whose horizon is located at $Z=\infty$\,. 

\medskip 

The Bekenstein-Hawking entropy can also be computed as 
\begin{eqnarray}
S_{\rm BH} = \left.\frac{A}{4 G_{\rm eff}}\right|_{Z\to \infty} 
= \left.\frac{\Phi^2}{4 G}\right|_{\sqrt{\mu}=\pi \, T_H}= \frac{1+\pi\,  T_{\rm H}}{4 G}\,.
\end{eqnarray}
Here the area $A$ is taken as $A=1$ because the horizon is just a point, and 
the effective Newton constant $G_{\rm eff}$ can be read off 
from the classical action as 
\begin{eqnarray}
\frac{1}{G_{\rm eff}} = \frac{\Phi^2}{G} \,. 
\label{Geff}
\end{eqnarray}

\medskip 

On the other hand, the holographic entropy can be computed by using the renormalized 
boundary stress tensor. For the detailed computation like the regularization and the counter term, 
see \cite{AP}. As a result, the renormalized boundary stress tensor is evaluated as
 \begin{eqnarray}
\langle \hat{T}_{tt} \rangle = \frac{\mu}{8\pi G} \equiv E\,. 
\end{eqnarray}
Then by using the thermodynamic relation 
\begin{eqnarray}
\dd S = \frac{\dd E}{T_{\rm H}}\,, \label{thermo}
\end{eqnarray}
the entropy is obtained as 
\begin{eqnarray}
S = \frac{\pi T_{\rm H}}{4 G} + S_{T_{\rm H}=0}\,,
\end{eqnarray}
where $S_{T_{\rm H}=0}$ is an integration constant. 
Thus the holographic entropy agrees with the Bekenstein-Hawking entropy,  
up to the temperature-independent constant. 

\medskip 

The main goal of this paper is to realize this correspondence of the entropies  
for a deformed black hole solution introduced in the next section.

\section{Deforming the AP model}

In this section, we consider deforming the AP model so that 
the deformed AdS$_2$ metric (\ref{3-metric}) is supported as a solution. 
For simplicity, the matter fields are turned off hereafter. Along this line, as well as the dilaton itself, 
the dilaton potential also has to be deformed from a simple quadratic one (\ref{AP}) 
to a hyperbolic function, similarly to integrable deformations. 

\subsection{The deformed AP model}

\subsubsection*{The deformed metric}

Before discussing the dilaton and the dilaton potential, 
it is helpful to rewrite the deformed metric (\ref{3-metric}) as 
\begin{eqnarray}
\dd s^2 &=& \frac{-\dd t^2 + \dd z^2}{z^2-\eta^2\left(\alpha + \beta t+\gamma (-t^2+z^2) \right)^2} 
\nonumber \\ 
&=& \frac{1}{1-\eta^2 (X\cdot P)^2}\frac{-\dd t^2+ \dd z^2}{z^2}\,. \label{d-met}
\end{eqnarray}
Here we have introduced new quantities:  a coordinate vector $X^I$ and 
a parameter vector $P_I$ defined as  
\begin{eqnarray}
X^I &\equiv& \frac{1}{z}\left(t~,~\frac{1}{2}(1+t^2-z^2)~,~\frac{1}{2}(1-t^2+z^2)\right)\,,\no\\
P_I &\equiv&(\beta~,~\alpha-\gamma~,~\alpha+\gamma)\,\qquad (I,J=1,2,3)\,.
 \end{eqnarray} 
The metric of the embedding space $\mathbb{M}^{2,1}$ is taken as $\eta_{IJ}=\text{diag}(-1,+1,-1)$\,. 
The inner products are defined as 
\begin{eqnarray}
&& X \cdot P \equiv X^I P_I = \frac{\alpha+\beta \, t+\gamma\,(-t^2+z^2)}{z}\,,  
 \label{def-para} 
\\ && X \cdot X \equiv \eta_{IJ}X^I X^J = -1\,, \qquad 
P\cdot P \equiv \eta^{IJ}P_I P_J = -\tilde{\omega}\,. 
\end{eqnarray}
These three products $X \cdot X$, $P \cdot P$ and $X\cdot P$ are transformed as scalars
under the $SL(2,\mathbb{R})$ transformation\footnote{
This $SL(2,\mathbb{R})$ transformation is the usual one generated by
three transformations, 
1) time translation, 2) dilatation and 3) special conformal transformation.}. 
For example, $X\cdot P$ is transformed as $X\cdot P= \tilde{X}\cdot \tilde{P}$, 
where $\tilde{X}$ and $\tilde{P}$ are new coordinate and parameter vectors, respectively. 

\medskip 

Using the $SL(2,\mathbb{R})$ transformation, we can choose the vector $\tilde{P}$ freely 
 as long as it satisfies the relation $\tilde{P}\cdot \tilde{P}=P\cdot P=-\tilde{\omega}$\,.
 Note that only the warped factor of the metric changes like 
 \begin{eqnarray}
 \dd s^2 &=&\frac{1}{1-\eta^2 (\tilde{X}\cdot \tilde{P})^2}
 \frac{-\dd\tilde{t}^2+\dd\tilde{z}^2}{\tilde{z}^2}\,
 \end{eqnarray}
because the rigid AdS$_2$ part is invariant under the $SL(2,\mathbb{R})$ transformation.

\subsubsection*{The dilaton sector}

Given the deformed metric (\ref{3-metric}) [or equivalently (\ref{d-met})]\,, 
by solving the equations of motion (\ref{2D-eom}) without the matter fields,
the dilaton $\Phi$ is determined as 
\begin{eqnarray}
\Phi^2 &=& \frac{c_1}{2\eta}\log\left|\frac{z+\eta\left(\alpha + \beta t 
+ \gamma (-t^2+z^2) \right)}{z-\eta\left(\alpha + \beta t 
+\gamma (-t^2+z^2) \right)}\right| +c_2 \nonumber \\ 
&\equiv & 
\frac{c_1}{2\eta}\log\left|\frac{1+\eta (X\cdot P)}{1-\eta (X\cdot P)}\right| +c_2\,, \label{deformed-dilaton}
\end{eqnarray}
when the dilaton potential is deformed as 
\begin{eqnarray}
U(\Phi)&=&
\left\{ \begin{array}{ll}\displaystyle 
~-(1-\tilde{\omega}\, \eta^2)\frac{c_1}{\eta}\sinh\left[\frac{2\eta}{c_1}(\Phi^2-c_2)\right] 
& ~~~\quad ({\text{for}}~~1>|\eta (X \cdot P)|)   
\vspace*{0.3cm} \\ \displaystyle
~+(1-\tilde{\omega}\, \eta^2)\frac{c_1}{\eta}\sinh\left[\frac{2\eta}{c_1}(\Phi^2-c_2)\right] 
& ~~~\quad  ({\text{for}}~~1<|\eta (X \cdot P)|)  \\
\end{array} \right.\,.
\label{d-pot}
\end{eqnarray}
Here $c_1$ and $c_2$ are arbitrary constants 

\medskip

In the undeformed limit $\eta \to 0$\,, the dilaton (\ref{deformed-dilaton}) is reduced to 
\[
\Phi^2 = c_2 + c_1\, \frac{\alpha + \beta\, t + \gamma (-t^2+z^2)}{z}\,,
\]
and thus the dilaton (\ref{dilaton-AP}) in the AP model has been reproduced 
when $c_1=1$ and $c_2 =1$\,.  Remarkably, the three parameters $\alpha$\,, $\beta$ and $\gamma$ 
correspond to $a$\,, $b$ and $c$ in (\ref{dilaton-AP})\,, respectively. 
Similarly, as $\eta \to 0$\,, the upper branch of the potential (\ref{d-pot}) reduces to 
\[
U(\Phi) = 2(c_2 - \Phi^2)\,,  
\]
while the lower branch vanishes. 
Thus the dilaton potential of the AP model is reproduced when $c_2 =1$\,. 
In total, the case with $c_1 = c_2 =1$ is associated with the AP model 
and hence we will work with $c_1=c_2=1$ hereafter. 

\subsubsection*{The vacuum solution in the deformed AP model}

In summary, the deformed AP model is specified by the deformed dilaton potential, 
\begin{eqnarray}
U(\Phi)&=&
\left\{ \begin{array}{ll}\displaystyle 
~-(1-\tilde{\omega}\, \eta^2)\frac{1}{\eta}\sinh\left[2\eta(\Phi^2-1)\right] 
& ~~~\quad ({\text{for}}~~1>|\eta (X \cdot P)|)   
\vspace*{0.3cm} \\ \displaystyle
~+(1-\tilde{\omega}\, \eta^2)\frac{1}{\eta}\sinh\left[2\eta(\Phi^2-1)\right] 
& ~~~\quad  ({\text{for}}~~1<|\eta (X \cdot P)|)  \\
\end{array} \right.\,, \nonumber 
\end{eqnarray}
and the vacuum solution is given by 
\begin{eqnarray}
 \dd s^2 = \frac{1}{1-\eta^2 (X\cdot P)^2}\frac{-\dd t^2+ \dd z^2}{z^2}\,, \qquad 
 \Phi^2 = \frac{1}{2\eta}\log\left|\frac{1+\eta (X\cdot P)}{1-\eta (X\cdot P)}\right| +1\,, 
\end{eqnarray}
where 
\begin{eqnarray}
 X \cdot P = \frac{\alpha+\beta \, t+\gamma\,(-t^2+z^2)}{z}\,. \nonumber 
\end{eqnarray}

\subsection{A deformed black hole solution}

In this subsection, we study a deformed black hole solution contained as a special case 
of the general vacuum solution (obtained in the previous subsection). 
This solution can be regarded as a deformation of the black hole solution presented 
in \cite{AP}. 

\medskip 

In the following,  instead of $\tilde{\omega}$\,, 
we use a new parameter $\mu$ defined as 
\[
\mu \equiv - \tilde{P}\cdot \tilde{P} =\tilde{\omega} = -4c\,,
\] 
so as to make our notation the same as that of the AP model. Here it may be worth noting 
that the black hole temperature is related to the modification of the CYBE. 
The zero temperature case corresponds to the homogeneous CYBE 
and the temperature is measured by negative values of $c$\,. Solutions of the mCYBE 
with negative (positive) $c$ are called the split (non-split) type. The well-known example 
of the non-split type is the $q$-deformation of AdS$_5$ \cite{DMV2}, while the split type 
has gotten little attention.  For the recent progress on the split type, see \cite{split}. 
It may be interesting to seek some connection between black hole geometries and 
solutions of split type. 

\medskip

By performing the same coordinate transformation as in the undeformed case like 
\begin{eqnarray}
x^\pm = \frac{1}{\sqrt{\mu}} \tanh(\sqrt{\mu}\, (T \pm Z))\,, 
\end{eqnarray}
the deformed black hole solution is obtained as 
\begin{eqnarray}
\dd s^2 &=&\frac{4\mu}{-\eta^2\mu+(1-\eta^2\mu)\sinh^2(2\sqrt{\mu}\, Z)}
\bigl(-\dd T^2+\dd Z^2 \bigr)\,,\no\\
\Phi^2&=&
1+\frac{1}{2\eta}{\text{log}}\left|
\frac{1+\eta\sqrt{\mu}\, \coth(\sqrt{\mu}\,Z)}{1-\eta\sqrt{\mu}\, 
\coth(\sqrt{\mu}\, Z)}\right|\,.
\label{d-BH}
\end{eqnarray}

\medskip 

\begin{figure}[htbp]
\begin{center}
\includegraphics[scale=0.3]{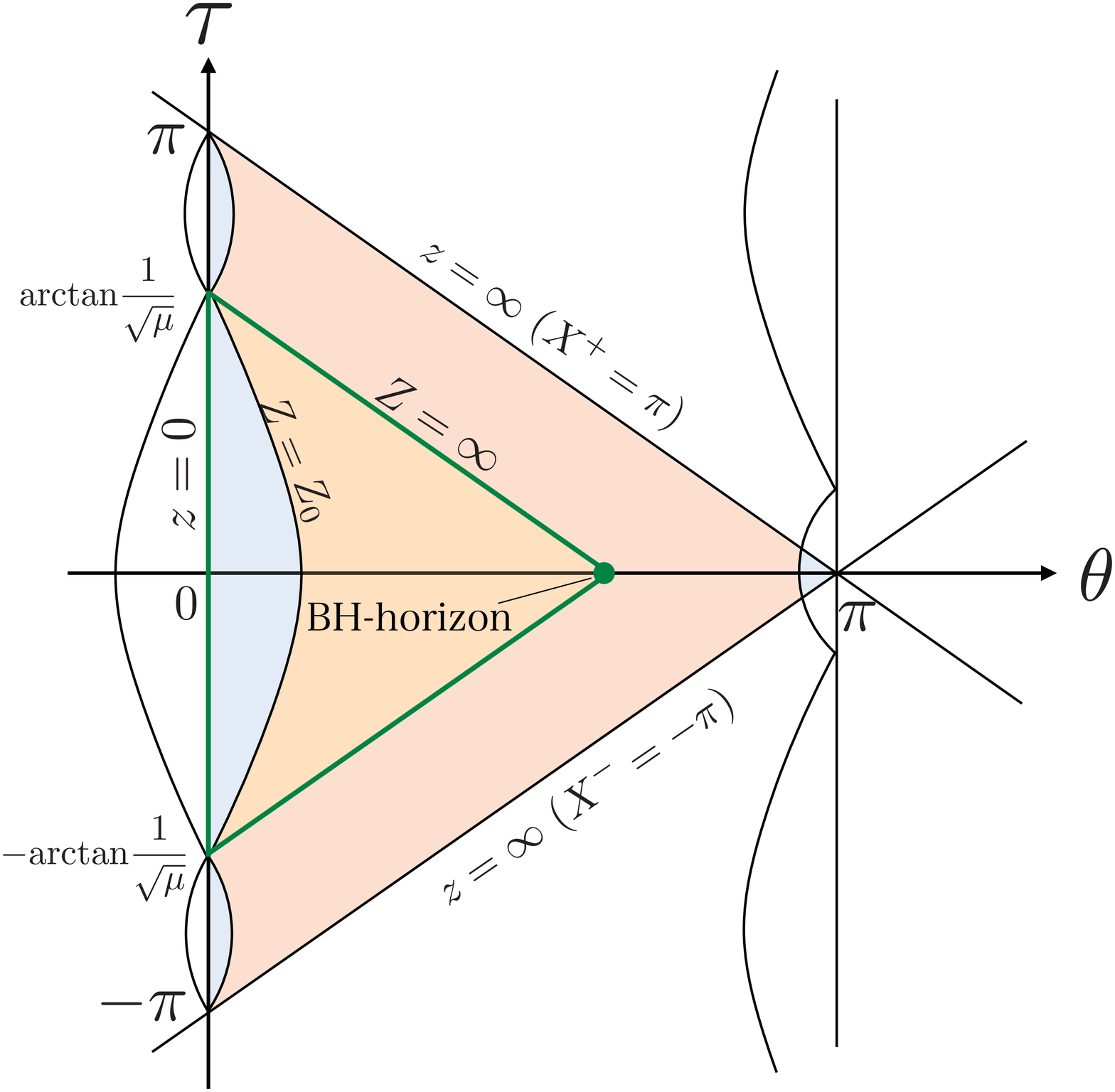}  
\vspace*{-0.5cm}
\caption{\label{fig:deformed AdS penrose} \footnotesize Penrose diagram of 
the deformed black hole. In this diagram, a curvature singularity is depicted 
in the global AdS$_2$ coordinates with $\alpha=1/2$, $\beta=0$ and $\gamma=\mu/2$ 
in (\ref{def-para}), where $\tau$ and $\theta$ are the same global coordinates as 
in the undeformed AdS$_2$\,. The black curves represent the curvature singularities 
of the deformed spacetime. In the blue region, the scalar curvature is positive, 
while in the red and orange regions, it takes negative values.
The black hole coordinates in (\ref{d-BH}) covers the interior bounded by the green lines.
By employing a Schwarzschild-like coordinate system (\ref{d-Sch})\,, 
we focus on the orange region in order to evaluate the black hole entropy.} 
\end{center}
\end{figure}

In this coordinate, the Ricci scalar (\ref{Ricci}) is rewritten as 
\begin{eqnarray}
R=-(1-\eta^2 \mu)\frac{1-\eta^2 \mu -(1+\eta^2 \mu)\text{cosh}(4\sqrt{\mu} Z) }{\eta^2 \mu-(1-\eta^2 \mu)\,\text{sinh}^2 (2\sqrt{\mu} Z) }\,.
\end{eqnarray}
In the following, we impose that 
\begin{eqnarray}
\eta^2 < \frac{1}{\mu} 
\label{condition}
\end{eqnarray}
so as to ensure the existence of  the undeformed limit\footnote{
Otherwise, it is not posible to take the undeformed limit 
$\eta\rightarrow 0$ because $\eta^2 >1/\mu$\,.}. 
Note here that this background has a naked singularity at $Z=Z_0$\,, where 
\begin{eqnarray}
Z_{0} \equiv \frac{1}{2\sqrt{\mu} }\, \text{arctanh}(\eta \sqrt{\mu})\,.  
\end{eqnarray}
This is a peculiar feature of the Yang-Baxter deformed geometry 
based on the modified CYBE like the $\eta$-deformation of AdS$_5$ \cite{ABF}.  
From (\ref{Ricci}), in the region with $Z>Z_0$ the Ricci scalar takes negative values, 
while for $0<Z<Z_0$, it has positive values (See Fig.\,\ref{fig:deformed AdS penrose})\,. 
In the undeformed limit $\eta \rightarrow 0$\,, $Z_0$ is sent to zero 
and the singularity disappears because the undeformed spacetime is just AdS$_2$\,.
In the following discussion, we focus upon the negative-curvature region ($Z>Z_0$)\,.
Therefore, we are concerned with only the upper branch of the potential (\ref{d-pot})\,. 

\medskip

By performing the following coordinate transformation, 
\begin{eqnarray}
r = \frac{1}{\eta} \text{arctanh}{ \bigl(\eta \sqrt{\mu}\, \coth \left(2 \sqrt{\mu}\, Z\right)
\bigr) }\,, \label{nice-coord}
\end{eqnarray}
the metric takes a Schwarzschild-like form\footnote{The factor 4 
is included so as to reproduce the result of \cite{AP}. }  
\begin{eqnarray}
\dd s^2 &=& -4 F(r)\,\dd T^2+\frac{ \dd r^2}{ F(r) }\,,  
\label{d-Sch}
\end{eqnarray}
where the scalar function $F(r)$ is defined as 
\begin{eqnarray}
F(r)&\equiv& \frac{-1-\eta^2 \mu +(1-\eta^2 \mu)\text{cosh}(2\, \eta \, r)}{2\eta^2}\,.
\label{func-F}
\end{eqnarray}
In this coordinate system, the dilaton takes the simplest form, 
\begin{eqnarray}
\Phi^2=1+r\,.
\end{eqnarray}
The locations of the boundary and black hole horizon are 
\begin{eqnarray}
\text{boundary}\,: ~~r= \infty\,,\qquad  \text{BH horizon}\,: ~~
r=r^*\equiv \frac{1}{\eta}\text{arctanh}(\eta\sqrt{\mu})\,.
\end{eqnarray}

\subsubsection*{Bekenstein-Hawking entropy}

Let us compute the Bekenstein-Hawking entropy of the deformed black hole 
by utilizing the coordinate system (\ref{d-Sch})\,. 

\medskip 

The Hawking temperature $T_{\rm H}$ is given by the standard formula: 
\begin{eqnarray}
T_{\rm H} &=& \left. \frac{1}{4 \pi} \partial_{r} \sqrt{ -\frac{ g_{tt} }{ g_{rr} } }\right|_{r=r^*}
= \frac{ \sqrt{\mu} }{\pi}\,. \label{HT}
\end{eqnarray}
This is the same result as the undeformed case. 
By assuming that the horizon area A is $1$ and using the effecting Newton constant 
$G_{\rm eff}$ in (\ref{Geff})\,, 
the Bekenstein-Hawking entropy $S_{\rm BH}$ can be computed as 
\begin{eqnarray}
S_{\rm BH} = \left. \frac{A}{4 G_{\rm eff}}\right|_{r=r^*} 
= \frac{\text{arctanh}{ (\pi \,T_{\rm H} \, \eta)} }{4 G \eta} + \frac{1}{4G}\,.
\label{BH}
\end{eqnarray}
In the undeformed limit $\eta \to 0$\,, the entropy is reduced to 
\[
S_{\rm BH}^{(\eta=0)} = \frac{\pi T_{\rm H}}{4G} + \frac{1}{4G}\,,
\]
and thus the result of AP model has been reproduced.

\section{The boundary computation of entropy}

In this section, we compute the entropy of the deformed black hole 
by evaluating the renormalized boundary stress tensor. 
Now that the boundary structure is drastically changed, the first thing is to 
determine the location of the holographic screen. In the following, we take the screen 
on the singularity by following the proposal of \cite{Kameyama}. More precisely, 
by introducing a UV cut-off $\epsilon$\,,  the boundary is taken just before 
the singularity ($Z=Z_0+\epsilon$)\,. 

\medskip 

In the conformal gauge, the total action including the Gibbons-Hawking term 
can be rewritten as
\begin{eqnarray}
S_{g,\Phi} &=&\frac{1}{16\pi G}\int\!\dd^2 x\, \sqrt{-g}\left[\Phi^2 R - U(\Phi)\right] +\frac{1}{ 8 \pi G} 
\int\!\dd t\, \sqrt{-\gamma}\,\Phi^2 K\,\no\\
&=&\frac{1}{8 \pi G}\int\!\dd^2 x \left[ 
-4 \partial_{(+} \Phi^2 \partial_{-)} \omega  - \frac{1}{2}U(\Phi)\, {\rm e}^{2 \omega} 
\right]\,.  
\end{eqnarray}
$K$ is the extrinsic curvature and $\gamma$ is the extrinsic metric. 
By using the explicit expression of the deformed black hole solution in (\ref{d-BH}), 
the on-shell bulk action can be evaluated on the boundary, 
\begin{eqnarray}
S_{g,\Phi} &=& \int \! \dd t\,  \left. \frac{ -\mu }{2\pi G ( 1+\eta^2 \mu +(-1+\eta^2 \mu )
 \cosh(4\mu^{ \frac{1}{2} } Z) ) } \right|_{Z\rightarrow Z_0}\,. 
\end{eqnarray}
Recall that the regulator $\epsilon$ is introduced such that $Z-Z_{0}=\epsilon$\,, 
the on-shell action can be expanded as 
\begin{eqnarray}
S_{g,\Phi} &=& \int \! \dd t\, \left[
\frac{1}{ 16 \pi G\, \eta\, \epsilon  } - \frac{1+\eta^2 \mu }{16 \pi G\, \eta^2}\,\epsilon^0 
+ \frac{ (3+\eta^2 \mu (-2+3\eta^2 \mu) )\,\epsilon }{48 \pi G\, \eta^3} +O(\epsilon^2) \right]\,. 
\end{eqnarray}

\medskip 

To cancel the divergence that occurs as the bulk action approaches the boundary, 
it is appropriate to add the following counter term:\footnote{The dual-theory interpretation 
of it is not so clear 
because it cotains an infinite number of polynomials and also depends 
on the temperature explicitly. Another counter term may be allowed and it would be nice 
to seek for it by following the procedure in \cite{Cvetic}. 
We are grateful to Ioannis Papadimitriou for this point. } 
\begin{eqnarray}
S_{\rm ct} &=& -\frac{1}{8 \pi G}\int\!\dd t\, \sqrt{-\gamma_{tt}}\, 
\sqrt{ F(\Phi^2-1) - \frac{1}{ \eta^2 } \log(1-\eta^2 \mu)}\,.  
\label{counter}
\end{eqnarray}
Here the scalar function $F$ is already given in (\ref{func-F}) and hence 
\begin{eqnarray}
 F(\Phi^2-1) &=& \frac{ -1-\eta^2 \mu +(1-\eta^2 \mu) \cosh( 2\eta (\Phi^2-1) )}{ 2 \eta^2 } 
\end{eqnarray}
Note that the inside of the root of (\ref{counter}) is positive 
due to the condition (\ref{condition})\,.
The extrinsic metric $\gamma_{tt}$ on the boundary is obtained as 
\[
\gamma_{tt} =\left. - {\rm e}^{2 \omega} \right|_{Z \to Z_{0}} \,. 
\]
In the undeformed limit $\eta \to 0$, this counter term reduces to
\begin{eqnarray}
S_{\rm ct}^{(\eta =0)} &=& \frac{1}{8 \pi G}\int\!\dd t\, \sqrt{-\gamma_{tt}}\, (1-\Phi^2)\,, 
\end{eqnarray}
because $\Phi^2-1>0$. This is nothing but the counter term utilized in the AP model \cite{AP}.

\medskip

It is straightforward to check that the sum $S = S_{g,\Phi}+S_{\rm ct}$ becomes finite 
on the boundary by using the expanded form of the counter term (\ref{counter}): 
\begin{eqnarray}
S_{\rm ct} &=& \int \! \dd t\, \left[
 \frac{-1}{16 \pi G \eta\, \epsilon } + \frac{1+\eta^2 \mu 
 +2 \log(1-\eta^2 \mu) }{ 16 \pi G\, \eta^2}\, \epsilon^0 +O(\epsilon) \right]\,.
\end{eqnarray}

\medskip

Around the boundary, the warped factor of the metric in (\ref{d-BH}) can be expanded as 
\begin{eqnarray}
 {\rm e}^{2 \omega} &=& \frac{1}{ \eta\, \epsilon } - \left[ \frac{1}{ \eta^2} +\mu \right]\,\epsilon^0
 + O(\epsilon)\,. 
\end{eqnarray}
Hence, by normalizing the boundary metric as 
\[
\hat{\gamma}_{tt}  = \eta\, \epsilon\, \gamma_{tt}\,, 
\]
the boundary stress tensor can be defined as 
\begin{eqnarray}
\langle \hat{T}_{tt} \rangle 
&\equiv& \frac{-2}{\sqrt{-\hat{\gamma}_{tt}}}\, \frac{\delta S}{\delta \hat{\gamma}^{tt} } 
= \lim_{ \epsilon \to 0 } \sqrt{\eta\, \epsilon}\, \frac{-2}{\sqrt{-\gamma_{tt} } }\, 
\frac{\delta S}{\delta \gamma^{tt} }\,.  
\end{eqnarray}
After all, $\langle \hat{T}_{tt} \rangle$ has been evaluated as  
\begin{eqnarray}
\langle \hat{T}_{tt} \rangle &=&-\frac{ \log(1-\eta^2 \mu) }{8 \pi G\, \eta^2}\,. 
\end{eqnarray}

\medskip

To compute the associated entropy, 
$\langle \hat{T}_{tt} \rangle$ should be identified with energy $E$ like 
\begin{eqnarray}
E &=& -\frac{ \log(1-\pi^2 T_{\rm H}^2 \eta^2 ) }{8 \pi G\, \eta^2}\,,
\end{eqnarray}
where we have used the expression of the Hawking temperature (\ref{HT})\,.  
Then by solving the thermodynamic relation (\ref{thermo}) again,  
the entropy is obtained as  
\begin{eqnarray}
S &=& \frac{ \text{arctanh}{(\pi T_{\rm H} \eta)} }{ 4 G \eta } + S_{T_{\rm H}=0}\,. 
\end{eqnarray}
Here $S_{T_{\rm H}=0}$ has appeared as an integration constant that measures 
the entropy at zero temperature. 
Thus the resulting entropy precisely agrees with the Bekenstein-Hawking entropy 
(\ref{BH})\,, up to the temperature-independent constant. 

\medskip 

Finally, it should be remarked that this agreement is quite non-trivial because 
the deformation changes the UV region of the geometry drastically. 
Hence the location of the holographic screen and the choice of the counter term 
are far from trivial. Although the holographic screen was supposed to be the singularity, 
inversely speaking, this agreement of the entropies here supports that the proposal 
in \cite{Kameyama} would work well. As for the geometrical meaning of the counter term 
(\ref{counter}), we have no definite idea. It is significant to figure out a systematic prescription 
to produce the counter term (\ref{counter})\,.

\section{Conclusion and discussion}

In this paper, we have discussed deformations of the AP model 
by following the Yang-Baxter deformation technique. To support the deformed 
AdS$_2$ metric, the dilaton itself is deformed and the dilaton potential is also modified 
from the polynomial to the hyperbolic function-type potential, similarly to integrable 
deformations. We have obtained the general vacuum solution for the deformed potential. 

\medskip 

A particularly interesting example is a deformed black hole solution. 
The deformation makes the spacetime structure 
around the boundary change drastically and a new naked singularity appears. 
The Hawking temperature is the same as in the undeformed case, 
but the Bekenstein-Hawking entropy is modified due to the deformation. 
This entropy has also been reproduced by evaluating the renormalized stress tensor 
with an appropriate counter term on the regularized screen close to the singularity.  

\medskip 

There are some open problems. A possible generalization is to include matter fields, 
though it has not succeeded yet. The matter contribution would not be so simple 
in comparison to the AP model. It is also interesting to consider lifting up our results to 
higher dimensional setups. Possibly, the most intriguing issue is to clarify the dual quantum mechanics 
for the deformed black hole presented here. A candidate would be a deformed 
SYK model which would be constructed by performing a disordered quench for a $q$-deformed 
Heisenberg magnet. When an infinitesimal deformation of the deformed AdS$_2$ geometry is considered, 
one would encounter a deformed Schwarzian derivative, though it seems difficult to determine 
what it is because there is no $SL(2)$ invariance on the boundary 
in comparison to the standard setup studied in \cite{Jensen,MSY,EMV}. It is also interesting to study 
Yang-Baxter deformations of the Callan-Giddings-Harvey-Strominger (CGHS) model \cite{CGHS}
by following \cite{BKLSY,MORSY}. 

\medskip  

There are some future directions associated with Yang-Baxter deformations as well. 
Now that we know the classical $r$-matrix which leads to the black hole geometry, 
it would be interesting to consider a Yang-Baxter deformation of higher-dimensional AdS 
with this $r$-matrix. In the study of Yang-Baxter deformations, it has been a long standing problem 
to determine where the holographic screen is, while there was a proposal for the $\eta$-deformed 
AdS$_5$ \cite{Kameyama} but it has not been supported by concrete evidence before this paper. 
It is significant to find out more supports to clarify the holographic interpretation 
for general Yang-Baxter deformations. 
 
\medskip 

We hope that the deformed AP model would provide a new arena to study 
the correspondence between nearly AdS$_2$ geometries and 1D quantum mechanical system 
like the SYK model or its cousins.

\subsection*{Acknowledgments}

We are very grateful to Ahmed Almheiri, 
Heng-Yu Chen, Koji Hashimoto, Keiju Murata, Ioannis Papadimitriou, 
Jun-ichi Sakamoto and Yuki Yokokura for useful comments and discussions.  
The work of H.K. is supported by the Japan Society for the Promotion of Science (JSPS).
The work of K.Y. is supported by the Supporting Program for Interaction-based Initiative Team Studies 
(SPIRITS) from Kyoto University and by a JSPS Grant-in-Aid for Scientific Research (C) No.\,15K05051.
This work is also supported in part by the JSPS Japan-Russia Research Cooperative Program 
and the JSPS Japan-Hungary Research Cooperative Program.

\end{document}